\documentstyle[12pt]{article}
\begin{document}
\date{\today}
\pagestyle{plain}
\title{Volkov solution for two laser beams and ITER}
\author{{\bf Miroslav Pardy}\\[2mm]
Laboratory of the Plasma Physics\\
and\\
Department of Physical Electronics \\
Masaryk University \\
Kotl\'{a}\v{r}sk\'{a} 2, 611 37 Brno, Czech Republic\\
e-mail:pamir@physics.muni.cz}
\date{\today}
\maketitle
\vspace{50mm}

\begin{abstract}
We find the solutions of the Dirac equation
for two plane waves (laser beams) and we determine the modified Compton formula
for the scattering of two photons on an electron.
The practical meaning of the two laser beams is, that two laser beams impinging
on a target which is constituted from material in
the form a foam, can replace 100-200 laser beams impinging on a normal
target and it means that the nuclear fusion with two laser beams is 
realistic in combination with the thermonuclear reactor such as ITER.

\end{abstract}

\newpage

\baselineskip 15 pt

\section{Introduction}

The application of a laser as a source of intense electromagnetic
radiation enables to study the new class of physical processes which are
running in the intense field of the electromagnetic wave.
The probability of some processes are increased, or, decreased
in comparison with the processes in vacuum. For
instance the probability of the process $\pi \rightarrow \mu + \nu$
is increased and the probability of the process $\pi \rightarrow e +
\nu$ is decreased. Some processes which are forbidden in vacuum, are allowed
in the intense field of the electromagnetic wave
($\nu \rightarrow \pi + \mu$, or, $e \rightarrow \pi + \nu$ ).
The polarization of the electromagnetic wave plays important role.

The situation with the two electromagnetic wave is the next
step and the future direction
of the laser physics of elementary particles. The two laser beams can
be used in the thermonuclear reactor instead of many laser beams.
Two laser beams impinging
on a target which is constituted from material in
the form of foam, can replace 100-200 laser beams impinging on a normal
target and it means that the nuclear fusion with two laser beams is 
realistic inside of the thermonuclear reactor such as ITER in
Cadarache near Aix-en-Provence in France. ITER means ``the way'' in
Latin and it will be prepared to produce 500 MW of fusion power in 2016 .

The laser field, or, the two laser fields is also the detector  
of the new properties of elementary particles
which cannot be revealed without laser. The situation with the
two independent lasers can give evidently further information on the
properties of elementary particles. At the same time the laser and the
system of two lasers can be
considered in chemistry as a specific catalyzer which was not known
before the existence of the laser physics. So, laser methods in
particle physics and chemistry can form in the near future the new
scientific revolution, which is not described in the standard
monographs on the scientific revolutions.

We consider here the electron described by the
the Dirac equation for two different four-potentials
of the plane electromagnetic
waves. We derive the partial differential equation for the wave
function, which is generalized form of the Volkov equation.
We find the solutions of the Dirac equation
for two orthogonal plane waves.
We determine the modified Compton formula for the scattering
of two photons on an electron.

The solution of the Dirac equation for the two waves was given by Sen
Gupta (Sen Gupta, 1967) in the form of the Fourier series, however
without immediate application.
The solution of the Dirac equation for the two waves
with the
perpendicular polarization was given for instance by Lyulka
(1974, 1975, 1977, 1985) who described the decay of particles in
the two laser fields.
The derivation of the two-wave solution is not explicitly involved
in Lyulka articles. So,we investigate the situation and present our results.

To be pedagogically clear,
we remind in the next section the derivation of the Volkov (1935)
solution of the Dirac equation in vacuum.

\section{Volkov solution of the Dirac equation with massless photons}

We follow the method of derivation and
metric convention of (Berestetzkii et al., 1989):

$$(\gamma(p-eA) - m)\psi = 0. \eqno(1)$$
where

$$A^{\mu} = A^{\mu}(\varphi); \quad \varphi = kx.\eqno(2)$$

We suppose that the four-potential satisfies the Lorentz gauge condition

$$\partial_{\mu}A^{\mu} = k_{\mu}\left(A^{\mu}\right)' =
\left(k_{\mu}A^{\mu}\right)' = 0, \eqno(3)$$
where the prime denotes derivative with regard to $\varphi$. From the
last equation follows

$$kA = const = 0,\eqno(4)$$
because we can put the constant to zero. The tensor of electromagnetic field is

$$F_{\mu\nu} = k_{\mu}A'_{\nu} - k_{\nu}A'_{\mu}.\eqno(5)$$

Instead of the linear Dirac equation (1), we consider the quadratic
equation, which we get by multiplication of the linear equation by
operator $(\gamma(p-eA) + m)$ (Berestetzkii et al., 1989).
We get:

$$\left[(p - eA)^{2} -m^{2} - \frac{i}{2}eF_{\mu\nu}
\sigma^{\mu\nu}\right]\psi = 0. \eqno(6)$$

Using $\partial_{\mu}(A^{\mu}\psi) = A^{\mu}\partial_{\mu}\psi$, which
follows from eq. (3), and $\partial_{\mu}\partial^{\mu} =
\partial^{2} = -p^{2}
$, with $p_{\mu} = i(\partial /\partial x^{\mu}) = i\partial_{\mu}$, we get
the quadratic Dirac equation for the four potential of the plane wave:

$$[-\partial^{2} - 2ie(A\partial) + e^{2}A^{2} - m^{2} -
ie(\gamma k)(\gamma A')]\psi = 0. \eqno(7)$$

We are looking for the solution of the last equation in the form:

$$\psi = e^{-ipx}F(\varphi).\eqno(8)$$

After insertion of eq. (8) into eq. (7), we get with ($k^{2} = 0$)

$$\partial^{\mu}F = k^{\mu}F', \quad \partial_{\mu}\partial^{\mu}F = k^{2}F''
= 0,\eqno(9)$$
the following equation for $F(\varphi)$

$$2i(kp)F' + [-2e(pA) + e^{2}A^{2} - ie(\gamma k)(\gamma A')]F = 0. \eqno(10)$$

The integral of the last equation is of the form (Berestetzkii et al., 1989):

$$F = \exp\left\{-i\int_{0}^{kx}\left[\frac {e(pA)}{(kp)} - \frac
{e^{2}}{2(kp)}A^{2}\right]
d\varphi + \frac {e(\gamma k)(\gamma A)}{2(kp)}\right\}
\frac{u}{\sqrt{2p_{0}}}, \eqno(11)$$
where $u/\sqrt{2p_{0}}$ is the arbitrary constant bispinor.

Al powers of $(\gamma k)(\gamma A)$ above the first are equal to zero,
since

$$(\gamma k)(\gamma A)(\gamma k)(\gamma A) =
- (\gamma k)(\gamma k)(\gamma A)(\gamma A) +
2(kA)(\gamma k)(\gamma A) = -k^{2}A^{2} = 0.\eqno(12)$$
where we have used eq. (4) and relation $k^{2} = 0$.
Then we can write:

$$\exp\left\{e\frac {(\gamma k)(\gamma A)}{2(kp)}\right\} =
1 + \frac {e(\gamma k)(\gamma A)}{2(kp)}.\eqno(13)$$

So, the  solution is of the form:

$$\psi_{p} = R \frac {u}{\sqrt{2p_{0}}}e^{iS}  =
\left[1 + \frac {e}{2kp}(\gamma k)(\gamma A)\right]
\frac {u}{\sqrt{2p_{0}}}e^{iS},
\eqno(14)$$
where $u$ is an electron bispinor of the corresponding Dirac equation

$$(\gamma p - m)u = 0\eqno(15)$$
and we shall take it to be normalized by condition ${\bar u}u = 2m$.
The mathematical object $S$ is the classical Hamilton-Jacobi function,
which  was determined in the form:

$$S = -px - \int_{0}^{kx}\frac {e}{(kp)}\left[(pA) - \frac {e}{2}
A^{2}\right]d\varphi. \eqno(16)$$

The current density is

$$j^{\mu} = {\bar \psi}_{p}\gamma^{\mu}\psi_{p},
\eqno(17)$$
where $\bar\Psi$ is defined as the transposition of (14), or,

$$\bar\psi_{p} = \frac {\bar u}{\sqrt{2p_{0}}}\left[1 +
\frac {e}{2kp}(\gamma A)(\gamma k)\right]
e^{-iS}.
\eqno(18)$$

After insertion of $\Psi_{p}$ and $\bar\Psi_{p}$
into the current density, we have:

$$j^{\mu} = \frac {1}{p_{0}}\left\{p^{\mu} - eA^{\mu} +
k^{\mu}\left(\frac {e(pA)}{(kp)} - \frac {e^{2}A^{2}}{2(kp)}\right)
\right\}.
\eqno(19)$$

\section{The solution of the Dirac equation for two plane waves.}

We suppose that the total vector potential is given as a superposition of the potential $A$ and $B$
as follows:

$$V_{\mu} = A_{\mu}(\varphi) + B_{\mu}(\chi),\eqno(20)$$
where $\varphi = kx$ and $\chi = \kappa x$ and $k \neq \kappa$.

We suppose that the Lorentz condition is valid. Or,

$$\partial_{\mu} V^{\mu} = 0 =
k_{\mu}\frac{\partial A^{\mu}}{\partial\varphi} +
\kappa_{\mu}\frac{\partial B^{\mu}}{\partial\chi} =
k_{\mu}A^{\mu}_{\varphi} + \kappa_{\mu}B^{\mu}_{\chi},\eqno(21)$$
where the subscripts $\varphi, \chi$ denote partial derivatives.
The equation (21) can be written in the more simple form
if we notice that partial differentiation with respect to
$\varphi$ concerns only $A$ and partial differentiation with respect to $\chi$
concerns only $B$. So we write instead eq. (21).

$$\partial_{\mu} V^{\mu} = 0 = k_{\mu}(A^{\mu})' + \kappa_{\mu}(B^{\mu})' = kA' + \kappa B'.\eqno(22)$$

Without loss of generality, we can write instead of equation (22) the following one

$$ k_{\mu}(A^{\mu})' = 0;  \quad \kappa_{\mu}(B^{\mu})' = 0; \quad {\rm or},\quad
kA = const = 0;\quad  \kappa B = const = 0, \eqno(23)$$
putting integrating constant to zero.

The electromagnetic tensor $F_{\mu\nu}$  is expressed in the new variables as in (5)

$$F_{\mu\nu}  = k_{\mu}A'_{\nu} - k_{\nu}A'_{\mu} +
\kappa_{\mu}B'_{\nu} - \kappa_{\nu}B'_{\mu}.\eqno(24)$$

Now, we can write Dirac equation for the two potentials the form

$$[-\partial^{2} - 2ie(V\partial) + e^{2}V^{2} - m^{2} -
\frac{i}{2}eF_{\mu\nu}\sigma^{\mu\nu}]\psi = 0. \eqno(25)$$
where $V = A + B$, $F_{\mu\nu}$ is given by eq. (24) and the combination of it with
$\sigma$ is defined as follows:

$$\frac{i}{2}eF_{\mu\nu}\sigma^{\mu\nu} = ie(\gamma k)(\gamma A') + ie(\gamma\kappa)(\gamma B')\eqno(26)$$

We will look for the solution in the standard Volkov form (8), or:

$$\psi = e^{-ipx}F(\varphi, \chi).\eqno(27)$$

After performing all operations prescribed in eq. (25), we get the
 following partial differential equation for the unknown function $F(\varphi,\chi)$:

$$ - 2k\kappa F_{\varphi\chi}  +  ( 2ipk - 2ik B)F_{\varphi} +
(2ip\kappa -2ieA\kappa) F_{\chi} \quad + $$

$$(e^{2}(A + B)^{2} - 2e(A+B)p  - ie(\gamma k)(\gamma A_{\varphi}) -
ie(\gamma \kappa)(\gamma B_{\chi}))F = 0.
\eqno(28)$$

\section{The solution of the Dirac equation for two orthogonal waves}

The equation (28) was simplified by author (Pardy, 2004b) putting 
$k\kappa = 0$. However, ignoring this simplification, we write  eq. 
(28) in the following form:

$$ a F_{\varphi} + b F_{\chi} + c F = 2k\kappa F_{\varphi\chi},\eqno(29)$$
where

$$a =  2ipk - 2iekB; \quad b =  2ip\kappa - 2ie\kappa A \eqno(30)$$
and

$$c = e^{2}(A + B)^{2} - 2e(A + B)p -
ie(\gamma k)(\gamma A') - ie(\gamma \kappa)(\gamma B').
\eqno(31)$$
and the term of with the two partial derivations
is not present because of $k\kappa = 0$.

For the field which we specify by the conditions

$$kB = 0; \quad \kappa A = 0; \quad AB = 0,\eqno(32)$$
we have:

$$2ipkF_{\varphi} + 2ip\kappa\kappa F_{\chi} + e^{2}A^{2} + 
e^{2}B^{2} -2epA -2epB -
ie(\gamma k)(\gamma A') \quad - $$

$$ie(\gamma \kappa)(\gamma B'))F = 2k\kappa F_{\varphi\chi}.\eqno(33)$$

Now, we are looking for the solution in the most simple form

$$F(\varphi,\chi) = X(\varphi)Y(\chi).\eqno(34)$$

After insertion of (34) into (33) and division the new
equation by $XY$ we get the terms
depending only on $\varphi$, and  on $\chi$. Or, in other words we
get:

$$\left(2i(pk + ik\kappa)\frac{X'}{X} +  e^{2}A^{2} - 2epA - 
ie(\gamma k)(\gamma A')\right)\quad + $$

$$\left(2i(p\kappa + ik\kappa)\frac{Y'}{Y} + e^{2}B^{2} -2epB  -
  ie(\gamma \kappa)(\gamma   B')\right) = 0\eqno(35)$$

So, there are terms dependent on $\varphi$  and terms dependent on 
$\chi$ only in eq. (35).
The only possibility is that they are equal to some
constant $\lambda$ and $-\lambda$. Then,

$$2i(pk + ik\kappa) X' +  (e^{2}A^{2} - 2epA - 
ie(\gamma k)(\gamma A'))X = \lambda X
\eqno(36)$$
and

$$2i(p\kappa + ik\kappa) Y' + (e^{2}B^{2} - 2epB - ie(\gamma \kappa)(\gamma B'))
Y = -\lambda Y\eqno(37)$$

We put $\lambda = 0$ without lost of generality.
Now, the solution of eq. (35) is reduced to the solution of
two equations only. Because the form of the equations is
similar to the form of  eq. (14) we can write the
solution of these equations as follows:

$$X = \left[1 + \frac {e}{2(kp + ik\kappa)}(\gamma k)(\gamma A)\right]
\frac {u}{\sqrt{2p_{0}}}e^{iS_1},
\eqno(38)$$
with

$$S_{1} =  \int_{0}^{kx}\frac {e}{(kp + ik\kappa)}\left[(pA) - \frac {e}{2}
(A)^{2}\right]d\varphi. \eqno(39)$$
and

$$Y = \left[1 + \frac {e}{2(\kappa p + ik\kappa)}(\gamma \kappa)(\gamma B)\right]
\frac {u}{\sqrt{2p_{0}}}e^{iS_2},
\eqno(40)$$
with

$$S_2 =  - \int_{0}^{\kappa x}\frac {e}{(\kappa p + ik\kappa)}\left[(pB) - \frac {e}{2}
(B)^{2}\right]d\chi. \eqno(41)$$

The total solution is then of the form:

$$\psi_{p} = \left[1 + \frac {e}{2(kp + ik\kappa)}(\gamma k)(\gamma A)\right]
\left[1 + \frac {e}{2(\kappa p + ik\kappa)}(\gamma \kappa)(\gamma B)\right]
\frac {u}{\sqrt{2p_{0}}}e^{i(S_{1}(A) + S_{2}(B))}.
\eqno(42)$$

\section{The standard Compton process}

In order to find the wave function of the electron in the two laser
beams, let us first remind the well known
Compton problem following from the Volkov solution. The
pioneering articles of this problem was written by Sen Gupta (1952)
solving the Compton scattering in the strong magnetic field. The next
application of the Volkov solution were done for the problem of pair
production by collision between a strong and coherent electromagnetic
field and a single energetic photon (Reiss, 1962). Later, Nikishov and
Ritus (Nikishov et al., 1964) and Goldman (1964a; 1964b) used the
Reiss ideas in their articles. 
Plenty of problems concerning the application of the Volkov solution
can be seen in the Ritus article (Ritus, 1979).

Let us consider electromagnetic monochromatic plane wave which is
polarized in a circle. We write the four-potential in the form:

$$A = a_{1}\cos\varphi + a_{2}\sin\varphi,\eqno(43)$$
where the amplitudes $a_{i}$ are equal in magnitude and orthogonal, or,

$$a_{1}^{2} = a_{2}^{2} = a^{2}, \quad a_{1}a_{2} = 0.\eqno(44)$$

Then, it possible to show that the Volkov solution for this situation is of
the form (Berestetzkii et al., 1989):

$$\psi_{p} = \left\{1 + \left(\frac {e}{2(kp)}\right)
[(\gamma k)(\gamma a_{1})\cos\varphi +
(\gamma k)(\gamma a_{2})\sin\varphi]\right\}\frac {u(p)}{\sqrt{2q_{0}}}
 \quad \times$$

$$\exp\left\{-ie\frac{(a_{1}p)}{(kp)}\sin\varphi + ie\frac
{(a_{2}p)}{(kp)}\cos\varphi -iqx\right\},\eqno(45)$$
where

$$q^{\mu} = p^{\mu} -e^{2}\frac {a^{2}}{2(kp)}k^{\mu}\eqno(46)$$
follows from eq. (19) as a time-average value. In other words, $q^{\mu}$ is
the mean value of quantity $p_{0}j^{\mu}$.

We know that the matrix element $M$ corresponding to the emission of
photon by an electron in the electromagnetic field is as follows
(Berestetzkii et al., 1989):

$$S_{fi} = -ie^{2}\int d^{4}x \bar \psi_{p'}(\gamma e'^{*})
\psi_{p}\frac {e^{ik'x}}{\sqrt{2\omega'}},
\eqno(47)$$
where $\psi_{p}$ is the wave function of an electron before interaction with
the laser photons and $\psi_{p'}$ is the wave function of electron after
emission of photon with components $k'^{\mu} = (\omega', {\bf k}')$.
The quantity $e'^{*}$ is the polarization four-vector of emitted photon.

The matrix element (47) involves the following linear combinations:

$$e^{-i\alpha_{1}\sin\varphi + i\alpha_{2}\cos\varphi}\eqno(48)$$

$$e^{-i\alpha_{1}\sin\varphi + i\alpha_{2}\cos\varphi}\cos\varphi\eqno(49)$$

$$e^{-i\alpha_{1}\sin\varphi + i\alpha_{2}\cos\varphi}\sin\varphi,\eqno(50)$$
where

$$\alpha_{1} = e\left(\frac {a_{1}p}{kp} - \frac {a_{1}p'}{kp'}\right),
\eqno(51)$$
and

$$\alpha_{2} = e\left(\frac {a_{2}p}{kp} - \frac {a_{2}p'}{kp'}\right),
\eqno(52)$$

Now, we can expand exponential function in the Fourier series, where
the coefficients of the expansion will be $B_{s}, B_{1s}, B_{2s}$. So we
write:

$$e^{-i\alpha_{1}\sin\varphi + i\alpha_{2}\cos\varphi}
= e^{-i\sqrt{\alpha_{1}^{2} + \alpha_{2}^{2}}\sin(\varphi - \varphi_{0})} =
\sum_{s = -\infty}^{\infty}B_{s}e^{-is\varphi}\eqno(53)$$

$$e^{-i\alpha_{1}\sin\varphi + i\alpha_{2}\cos\varphi}\cos\varphi
= e^{-i\sqrt{\alpha_{1}^{2} + \alpha_{2}^{2}}\sin(\varphi -
  \varphi_{0})}
\cos\varphi = \sum_{s = -\infty}^{\infty}B_{1s}e^{-is\varphi}\eqno(54)$$

$$e^{-i\alpha_{1}\sin\varphi + i\alpha_{2}\cos\varphi}\sin\varphi
= e^{-i\sqrt{\alpha_{1}^{2} + \alpha_{2}^{2}}\sin(\varphi -
\varphi_{0})}\sin\varphi =
\sum_{s = -\infty}^{\infty}B_{2s}e^{-is\varphi}\eqno(55)$$

The Coefficients $B_{s}, B_{1s}, B_{2s}$ can be expressed by means of
the Bessel function as follows (Berestetzkii et al., 1989):

$$B_{s} = J_{s}(z)e^{is\varphi_{0}}\eqno(56)$$

$$B_{1s} = \frac {1}{2}\left[J_{s+1}(z)e^{i(s+1)\varphi_{0}} +
J_{s-1}(z)e^{i(s-1)\varphi_{0}}\right]\eqno(57)$$

$$B_{2s} = \frac {1}{2i}\left[J_{s+1}(z)e^{i(s+1)\varphi_{0}} -
J_{s-1}(z)e^{i(s-1)\varphi_{0}}\right],\eqno(58)$$
where the quantity $z$ are now defined through the $\alpha$-components, or,

$$z = \sqrt{\alpha_{1}^{2} + \alpha_{2}^{2}}\eqno(59)$$
and

$$\cos\varphi_{0} = \frac {\alpha_{1}}{z}; \quad \sin\varphi_{0} =
\frac{\alpha_{2}}{z}.\eqno(60)$$

Functions $B_{s}, B_{1s}, B_{2s}$ are related one to another as follows:

$$\alpha_{1}B_{1s} + \alpha_{2}B_{2s} = sB_{s},\eqno(61)$$
which follows from the well known relation for Bessel functions:

$$J_{s-1}(z) + J_{s+1}(z) = \frac {2s}{z}J_{s}(z).\eqno(62)$$

The matrix element (47) can be written in the form (Berestetzkii et al., 1989):

$$S_{fi} = \frac {1}{\sqrt{2\omega'2q_{0}2q_{0}'}}\sum_{s}M_{fi}^{(s)}
(2\pi)^{4}i\delta^{(4)}(sk + q - q'- k'),\eqno(63)$$
where the $\delta$-function involves the law of conservation:

$$sk + q = q' + k'; \quad s = 1, 2, 3, ...\eqno(64)$$
with the relation

$$q^{2} = q'^{2} =
m^{2}(1 + \xi^{2}) \equiv m_{*}^{2} = m^{2}\left(1 -
\frac{e^{2}a^{2}}{m^{2}}\right),\eqno(65)$$
as it follows from eq. (46).

For  $s= 1$, eq. (64) has the physical  meaning of the conservation of energy-
momentum of the  one-photon  Compton process,
$s = 2$ has meaning of the two-photon Compton process and $s = n$
has meaning of the multiphoton interaction with $n$ photons.

It is possible to show, that the differential probability
per unit volume and unit time of the emission of the $s$ harmonics is of the
following form (Berestetzkii et al., 1989):

$$dW_{s}= |M_{fi}^{(s)}|^{2}\frac {d^{3}k' d^{3}q'}{(2\pi)^{6}2\omega' 2q_{0}
2q_{0}'}(2\pi)^{4}\delta^{(4)}(sk + q - q' - k').\eqno(66)$$

In order to obtain the probability of emission of photon, we must make some
operation with the matrix element $M$.
We will here not perform these operations.
We concentrate our attention on the conservation law in the formula (66).
It can be expressed by words as follows. The multiphoton object with the momentum $sk$ interacts with the electron of
the momentum $q$, and the result is the electron with the momentum $q'$ and one photon with the momentum $k'$.

Now, let us consider the equation (64) in the form

$$sk + q - k' = q'.\eqno(67)$$

If we introduce the angle $\Theta$ between ${\bf k}$ and ${\bf k}'$, then,
with  $|{\bf k}| = \omega$  and  $|{\bf k}'| = \omega'$,  we get from
the squared equation (67) in the rest system of electron, where
$q = (m_{*},0)$, the following equation:

$$s\frac {1}{\omega'} - \frac {1}{\omega} = \frac {s}{m_{*}}(1 - \cos\Theta),
\eqno(68)$$
which is modification of the original equation for the Compton process

$$\frac {1}{\omega'} - \frac {1}{\omega} = \frac {1}{m}(1 - \cos\Theta).
\eqno(69)$$

So, we see that Compton effect described by the Volkov solution of
 the Dirac equation
differs from the original Compton formula only by the existence of the
renormalized mass and  parameter $s$ of the multiphoton interaction.

We know that the last formula of the original Compton effect can be written
in the form suitable for the experimental verification, namely:

$$\lambda' - \lambda = \Delta \lambda = 4\pi\frac{\hbar}{mc}\sin^{2}\frac {\Theta}{2},
\eqno(70)$$
which was used by Compton for the verification of the quantum
nature of light.

\section{The two-photon Compton process}

In case of the two laser beams which are not collinear
the experimental situation involves possibility that the two different
photons can interact with one electron. The theory does not follow
from the standard one-photon Volkov solution because in the standard
approach the multiphoton interaction involve the collinear photon s and
not photons from the two different lasers. The problem was solved by
Lyulka in 1974 for the case of the two linearly polarized waves (Lyulka, 1974)

$$A = a_{1}\cos\varphi; \quad B = a_{2}\cos(\chi + \delta) \eqno(71)$$
with the standard conditions for $\varphi, \chi, k, \kappa$. The
quantity $\delta$ is the phase shift.

The two-wave Volkov solution is given by eq. (42) and the matrix
elements and appropriate ingredients of calculations are given by the
standard approach as it was shown by Lyulka (Lyulka, 1974).

It was shown (Lyulka, 1974), that

$$q^{\mu} = p^{\mu} - e^{2}\frac {a_{1}^{2}}{2(kp)}k^{\mu} -
e^{2}\frac {a_{2}^{2}}{2(\kappa p)}\kappa^{\mu}
\eqno(72)$$
and

$$ m_{*}^{2} = m^{2}\left(1 -
\frac{e^{2}a_{1}^{2}}{m^{2}} - \frac{e^{2}a_{2}^{2}}{m^{2}}\right).
\eqno(73)$$

The matrix element involves
the extended law of conservation. Namely:

$$sk + t\kappa  + q  =  q' +  k' + \kappa',\eqno(74)$$
where $s$ and $t$ are natural numbers
and the interpretation of the last equation is evident.
The multiphoton objects with momenta  $sk$ and $t\kappa$
interact with electron with momentum $q$. After interaction the
electron has a momentum $q'$
and two photons are emitted with
momenta $k'$ and $\kappa'$.

Instead of equation (74), we can write

$$sk + q - k' = q' + \kappa'  - t\kappa.\eqno(75)$$

From the squared form of the last equation and after some modification,
we get the following generalized equation of the double Compton
process for $s = t = 1$:

$$\frac {1}{\omega'} - \frac {1}{\omega} = \frac {1}{m_{*}}(1 -
\cos\Theta) + \frac{\Omega' - \Omega}{\omega\omega'} - \frac{\Omega\Omega'}{\omega\omega' m_{*}}(1 - \cos \Xi),
\eqno(76)$$
where the angle $\Xi$ is the angle between the 3-momentum of the
$\kappa$-photon  and the 3-momentum of the $\kappa'$-photon with
frequency $\Omega$ and $\Omega'$ respectively.

Let us remark that if the frequencies of the photons of the first wave substantially
differs from  the frequency of photons of the second electromagnetic
wave, then, the derived formula (76) can be experimentally verified by
the same way as the original Compton formula. To our knowledge, formula
(76) is not involved in the standard textbooks on quantum
electrodynamics because the two laser physics is at present time not
sufficiently developed.

\section{Discussion}

We have discussed  the problem of the Dirac equation with the two-wave
potentials of the electromagnetic fields. While the Volkov solution
for one potential is well known for long time, the case with the two waves
represents the new problem.
To our knowledge, the Compton process with two beams was not
investigated experimentally by any laboratory.

This article is in a some sense author's continuation of the problems
where the Compton, or
Volkov solution plays substantial role (Pardy, 1998; 2001; 2003;
2004a; 2004b).

It is possible to consider the situation with sum of  N waves, or,

$$V = \sum_{i = 1}^{N}A_{i}(\varphi_{i})\quad \varphi_{i} = k_{i}x.\eqno(77)$$

The problem has obviously physical meaning because the problem of the
laser compression of target by many beams
is one of the prestigious problems of the today laser physics. The
goal of the experiments is to generate the physical process of
implosion. When an intense petawat laser light is uniformly impinged on a
spherical fuel pellet,the   
laser energy is absorbed on the surface to generate a high-temperature
plasma of 2-3 keV and an extremely high pressure of a few hundred
megabars is generated. The pressure accelerates the outer shall of the
target towards the target centre. If the dynamics is sufficiently   
spherically symmetric, the central area is heated  up to 5-10 keV 
and fusion reaction starts (Nakai and Mima, 2004). 
The solution of that problem
in the general  form is not elementary  and can be solved only
by some laser institution such as the Lebedev
institution of physics, the  Livermore laser national laboratory and
so on.

On the other hand, if the target is constituted from material in
the form a foam, then instead of using 100-200 laser beams it is
possible to use only two laser beams, as it is supposed in our
article. Then the nuclear fusion is more realistic in the situation of
the thermonuclear reactor (Rozanov, 2004).
 
Nuclear fusion involves the bringing together of atomic nuclei.
The sum of the individual masses of the nucleons is greater than the
mass of the whole nucleus. This is because the strong nuclear force
holds the nucleons together. Then, the combined nucleus is a lower
energy state than the nucleons separately. The energy difference is
released in the fusion process.  

There are two major fusion processes. The magnetic confinement and
inertial confinement. The inertial fusion occurs inside targed fuel
pellets by imploding them with laser or particle beam irradiation in
brief pulses. It produces extremely high densities in the targed where
the laser pulse creates a schock wave in the pellet that it
intensified by its internal geometry. On the other hand, magnetic
fusion devices, like the tokamak, operate at lower densities, but use
magnetic fields to confine the plasma for longer time. 

To achieve a burning plasma, a sufficiently high density of 
fuel must be heated to temperatures about 100 million degrees of
Celsius that the nuclei collide often enough despite their natural
repulsive forces and energy losses. 

The fuels  to be used are two isotopes of hydrogen. Namely, deuterium and
tritium. While deuterium occurs naturally in sea water and it means it
is inexhaustible, tritium can be
bred in a fusion system when the light element, lithium, absorbs
neutrons produced in the fusion reaction. World resources of lithium
are inexhaustible and it means that also the energy obtained by fusion
process is practically infinite.  

\vspace{20mm}

\noindent
{\bf References}

\vspace{10mm}

\noindent
Berestetzkii, V. B., Lifshitz,  E. M.  and  Pitaevskii, L. P. (1989).
{\it Quantum Electrodynamics}, Moscow, Nauka, (in Russian). \\[2mm]
Goldman, I. I. (1964a). Intensity effects in Compton scattering,
{\it Zh. Exp. Teor. Fiz.} {\bf 46}, 1412;
ibid. (1964b). Intensity effects in Compton scattering,
{\it Phys. Lett.} {\bf 8}(2), 103. \\[2mm]
Lyulka, V. A. (1974).  Quantum effects in  intense
electromagnetic fields, {\it Sov. Phys. JETP }{\bf 40}, 815.\\[2mm]
Lyulka, V. A. (1975). {Decays of elementary particles in the field of
of the intense electromagnetic wave},  {\it Zh. Exp. Teor. Fiz.} {\bf 69}, Vol. 3(9), 800.\\[2mm]
Lyulka, V. A. (1977). {Quantum effects in the field of
the nonchromatical electromagnetic wave}, {\it Zh. Exp. Teor. Fiz.} {\bf 72}, Vol. 3,  865.\\[2mm]
Lyulka, V. A. (1985). {Weak processes in the intense Electron in 
the ultrashort laser pulse
electromagnetic fields}, Journal of Nuclear Physics Vol. 5(11), 1211. \\[2mm]
Nakai, S. and Mima, K. (2004). Laser driven inertial fusion energy:
present and perspective, {\it Rep. Prog. Phys.} {\bf 67}, 321.\\[2mm]  
Nikishov, A. I. and Ritus, V. I. (1964). Quantum processes in the
field of electromagnetic wave in the constant field I., 
{\it Zh. Eksp. Teor. Fiz.} {\bf   46}, 776.\\[2mm] 
Pardy, M. (1998). {The quantum field theory of laser
  acceleration}, Physics  Letters {\bf A 243}, 223. \\[2mm]
Pardy, M. (2001). The quantum electrodynamics of laser acceleration,
{\it Radiation Physics and Chemistry} {\bf 61}, 321.\\[2mm]
Pardy, M. (2003). {Electron in the ultrashort laser pulse}, 
{\it International Journal of Theoretical Physics} {\bf 42}(1), 99.\\[2mm]
Pardy, M. (2004a). Massive photons and the Volkov solution,
{\it International Journal of Theoretical Physics}
{\bf 43}(1), 127.\\[2mm]
Pardy, M. (2004b). Volkov solution for an electron in two wave
fields, e-print hep-ph/0408288.\\[2mm]
Reiss, H. R. (1962). Absorption of light by light, {\it
  J. Math. Phys.} {\bf 3}, 59.\\[2mm]
Rozanov, V. B. (2004). On the possible realization of spherical
compression for fusion targets irradiated by two laser beams, {\it Uspekhi
Fiz. Nauk}, {\bf 174}(4), 371. (in Russian).\\[2mm]
Ritus, V. I. (1979). {The quantum effects of the interaction of
elementary particles with the intense electromagnetic field}, 
{\it Trudy FIAN} {\bf 111}, 5.\\[2mm]
Sen Gupta, N. D. (1952). {\it Bull. Calcutta Math. Soc.}, {\bf 39}, 147;
ibid. {\bf 44}, 175. \\[2mm]
Sen Gupta, N. D. (1967). On the solution of the Dirac equation in the
field of two beams of electromagnetic radiation, {\it Zeitschrift 
f$\ddot{u}$r Physik} {\bf 200}, 13.\\[2mm]
Volkov, D. M. (1935). $\ddot{\rm U}$ber
eine Klasse von L$\ddot{\rm o}$sungen der Diracschen Gleichung,
{\it Zeitschrift f$\ddot{u}$r Physik} {\bf 94}, 250.

\end{document}